\documentclass[11pt]{article}
\usepackage{epsf}
\usepackage{graphicx}        

\def\ll{\label}

\def\c{\cite}

\def\r1{(\ref{$1})}

\def\ba{\begin{array}{c}}

\def\ea{\end{array}}

\def\l{\left}
\def\l({\left(}
\def\r){\right)}
\def\r{\right}

\def\la{\lambda}

 \def\be{\begin{equation}}
\def\bc{\begin{center}}
\def\ec{\end{center}}
\def\bit{\begin{itemize}}
\def\eit{\end{itemize}}
\def\ee{\end{equation}}
\def\ed{\end{document}}
\def\bea{\begin{eqnarray}}
\def\eea{\end{eqnarray}}
\def\efr{\end{flushright}}

\begin{document}
\title{{ Hidden possibilities in controlling optical soliton in 
fiber guided 
doped resonant  medium 
}}
\author{ Anjan Kundu\\ 
 Theory Group \& CAMCS, Saha Institute of Nuclear Physics,  
 Calcutta, INDIA}
\maketitle



 {\small Fiber guided optical signal propagating in a Erbium doped
nonlinear resonant medium is known to produce cleaner solitonic pulse,
described by the  self induced transparency (SIT) coupled to nonlinear
 Schr\"odinger  equation.  We discover two new possibilities hidden
in its integrable structure, for  amplification and  control of 
the optical pulse. Using the variable soliton width 
 permitted by the integrability of this model, the broadening   pulse
  can be regulated by
adjusting the initial population inversion  of the
dopant atoms.  The effect can be enhanced by another innovative application of
 its  constrained integrable  hierarchy, proposing a
 system of   multiple SIT media.  These
 theoretical predictions are workable analytically in details, correcting 
 a well known result.
}
\medskip

\noindent {\bf I. INTRODUCTION}

Optical communication through fiber has achieved phenomenal development
over the last two decades \c{OptCom}.  
 Dissipation and  dispersion in the
media, which are the  main hindrances in  signal
transmission, 
are usually attempted  to be solved 
by the dispersion management techniques and  devices \c{OptCom,dismang}.
 On the other hand, 
 in soliton based  optical communication, 
mediated by the  
nonlinear Schr\"odinger  
(NLS) equation
proposed much earlier \c{nlssolit}, 
 the group velocity dispersion can be countered
by  the
self phase modulation  in the 
nonlinear fiber medium \c{agarwal}. However,  the
  experiments revealed  
insufficiency of the model for its efficient practical application
\c{nlsdrab}. Another   proposal
 with improved 
 solitonic transmission was   due to the
 self-induced  transparency 
(SIT), 
 produced  by the 
  coherent response of the medium to an ultra short
 optical pulse  \c{SIT,lamb}.  Finally,
the benefits of both the   NLS and the  SIT systems,
were   combined  in a coupled NLS-SIT model \c{Maimitsov},
  by  transmitting the optical soliton 
through an Erbium 
 doped  nonlinear resonant medium
  \c{nakazawa1,nakazawa2}.

 However   the   solitonic communication,  
 in spite of  its favorable features and   
 theoretical  
advantages due to   its underlying
  integrability, 
  did not receive the
 needed response. 
 Our  aim here therefore, is 
to revisit   the NLS-SIT model for exploring
  new  possibilities   hidden in its integrable structures
and use them for the  control and amplification of the  solitonic pulse.
Though the  soliton usually moves with a constant velocity or speed, 
the integrable property of this coupled NLS-SIT system, as we find here,
   allows 
the soliton speed to be a tunable   function. And  since 
  the soliton  width in this case is  related to its speed, 
which in turn is   linked  here to the initial 
  population inversion, the  pulse width 
can be regulated  by manipulating the  population inversion profile
of the dopant atoms. This controlling effect can be   further enhanced
by exploring  another   specialty  of this integrable system, namely
its constrained   integrable  hierarchy, through a novel
 use of   coupled multiple   SIT media, in place of the conventional
single doped medium. 
 The  details can be  worked out  
 exactly  due to the integrability of the model,
  detecting the limitation of a well known
result  on   NLS-SIT soliton \c{nakazawa1,nakazawa2}.

Propagation of a stable optical pulse through a fiber medium,
 serving as a dispersive and nonlinear wave guide 
with Kerr nonlinearity \c{nlssolit},  can
 be described by the optical electric field $E(z,t) $ satisfying
 the well known  NLS equation \ 
\be \ i E_z-E_{tt}- 2 |E |^2 E=0, \ 
\ll{nls} \ee
 with space and time variables being  interchanged
as customary in nonlinear optics \c{agarwal}.
On the other hand,
   ultra short
 optical pulse   producing   SIT 
  in  the medium can be described  by the
 Maxwell-Bloch equation \c{SIT,lamb}
\be i E_z= 2p, \ \ 
ip_{t}=2 N E, \ iN_{t}=-(E p^*-E^* p), \ll{SIT} \ee
with the  induced polarization $p $
and the population inversion $N $ of the medium, contributed by the
Bloch equation. 
Fascinatingly, it is possible to combine these two effects
by transmitting the stable nonlinear pulse produced in the fiber 
wave guide through a doped medium with coherent response, governed
by  a  coupled NLS-SIT system given by 
a deformed  NLS equation
\be i E_z-E_{tt}- 2 |E |^2 E = 2p_1, \ \ll{pnls} \ee
  together with the   SIT  equations
  \be
ip_{1t}=2 (N_1 E-w_0 p_1), \ iN_{1t}=-(E p_1^*-E^* p_1), \ll{pN} \ee
representing  a nonholonomic constraint \c{kundu09}. 
 In (\ref{pN}) ,
  $ \  p_1=\nu \tilde \nu^{*}  $ 
is the  induced polarization   and  
  $N_1=|\tilde \nu |^2-|\nu |^2, \ -1 \leq N_1 \leq 1 , \ $ is the 
 population inversion 
 of the two level dopant  atoms
with  normalized wave functions 
 $\nu , \ \tilde \nu $ for the  ground  and  the excited
states,
respectively and  $ w_0$ is the natural frequency of these resonant ions.
 Assuming a homogeneous  broadening 
 of the
frequency spread with a  sharp resonance at $\Delta w= w-w_0 $,
 we have taken the symmetric distribution $ g(\Delta w)=\delta(\Delta w) $
 and replaced the average value $<p_1>=\int d w g(\Delta w)
p_1(z,t,w) $ appearing in (\ref{pnls}) by $p_1=p_1(z,t,w_0) $ and normalized the
coupling constants to ensure the integrability of the model.
\\ \\
{\bf II.  INTEGRABILITY AND SOLITON SOLUTION}


Recall that,  an integrable nonlinear equation may be  associated with a 
linear system $\Phi_t =U \Phi, \ \Phi_z = V \Phi, $
defined through a Lax pair   $U(\la),V(\la) $,  which  
are matrices with  their elements depending on the basic fields and a
parameter $\la $, known as the spectral parameter.
Through compatibility of the linear Lax equations, inducing 
 flatness condition: $U_z-V_t +[U,V]=0, $ the Lax pair 
 yield   the given nonlinear equation and   
 at the same time can be used 
to extract its  exact  solutions  through the 
 inverse scattering method (ISM)  \c{soliton}.  
We remind again that,  the    space $z $ and time $t$
are  interchanged here  in the context of nonlinear optics.
  It is noteworthy that, while the set of
 coupled NLS-SIT equations (\ref{pnls}-\ref{pN})
generalize both the NLS (\ref{nls}) and the  SIT (\ref{SIT}) equations,  
the   associated   
 Lax pair $U_{nls:sit}, V_{nls:sit} $  contain these subsystems    as the
constituent parts:
\be U_{nls:sit}(\la)=U_{nls}(\la )=U_{sit}(\la ), \
\ \ V_{nls:sit}(\la)=V_{nls}(\la)+ V_{sit}(\la) \ll{UV}\ee
where $U_{nls}, V_{nls} $ and
 $U_{sit}, V_{sit} $  are the  Lax pairs related to the 
 NLS and the SIT equations, respectively. The  NLS Lax pair is  well known as
 \c{soliton}
\bea U_{nls}(\la)&=&i (\sigma^3 \lambda + U^{(0)}), \
U^{(0)}= E\sigma^+ + E^*\sigma^-  
 \ll{Unls}\\
 V_{nls}(\la)&=&V^{(0)}+V^{(1)}\la +
V^{(2)}\la^2, \ \nonumber \\
  V^{(0)}&=& (U_{x}^{(0)}-i(U^{(0)})^2) \sigma^3, \
 V^{(1)}=2iU^{(0)}, \ V^{(2)}= 2i\sigma^3, \ll{Vnls} 
\eea
where $\sigma^3, \ \sigma^\pm=\frac 1 2 (\sigma^1 \pm i \sigma^2) $
 are the  $2\times 2 $ Pauli matrices. 
The SIT Lax pair can be given by the same  
 {\it time}-Lax operator as that of the NLS: $U_{sit}(\la)=U_{nls}(\la), $
while  the 
{\it space}-Lax operator  
\be V_{sit}(\la)=i     (\la - w_{0})^{-1} G_{1}, \
 G_{1}= N_1 \sigma^3+p_1 \sigma^+ + p^*_1 \sigma^- , \ll{Vsit}\ee
can be linked to the nonholonomic deformation \c{kundu09}. 
 We  check easily that, the flatness condition of 
 (\ref{Unls},\ref{Vnls})  
yields  the NLS equation (\ref{nls}), while  (\ref{Unls},\ref{Vsit})
 the SIT equation (\ref{SIT}) and similarly, the Lax pair  (\ref{UV})
would yield the coupled NLS-SIT equations (\ref{pnls},\ref{pN})

The {\it time}-Lax operator $U(\la ) $ plays the central role
in  the ISM  for finding the exact solutions of the nonlinear equation 
\c{soliton} and therefore, since   $U(\la ) $
 is the same for NLS (\ref{nls}), SIT (\ref{SIT}) and NLS-SIT
(\ref{pnls},\ref{pN}) equations as seen from (\ref{UV}),
  the form of  soliton solutions and the ISM
procedure  are remarkably similar for  all the three 
equations.
We therefore  present   soliton solutions for all of them
in an unified way following the ISM \c{soliton}, which
 though  an involved method, gives the 1-soliton solution in an amazingly
 simple  form:
\be E= c \ \frac  g {1+|g |^2}, \ \  g= {\rm exp} [2i({\rm u}^\infty 
t+ \tilde{\rm v}^\infty
z+\phi)]
, \ll{Esol1}\ee
 with $c, \phi =$ constants.  Note that,  the crucial elements 
${\rm u}^\infty, \tilde {\rm v}^\infty =
\frac 1 z \int^z dz {\rm  v}^\infty  $ in  
 (\ref{Esol1}), though linked to the Lax pair of the given system,
  need information only about their asymptotic properties:
$ {\rm u}^\infty= \sigma^3
U(\la _1)|_{t = -\infty}, \ 
{\rm v}^\infty= \sigma^3
V(\la _1)|_{t = -\infty} ,$
at  discrete spectral parameter  $\la _1 $.
 Therefore, fixing the 
initial condition of the basic fields involved in the NLS-SIT equations as
\be  E(z,t = -\infty) \to 0, \ 
p_1(z,t = -\infty) \to 0,   N_1(z,t = -\infty)= N_0(z) 
 \ll{BC}\ee
with an arbitrary function $ N_0(z), $ we can easily derive
from the Lax pair
(\ref {UV}-\ref {Vsit}):
 \bea  \sigma^3 {\rm u}_{nls:sit}^\infty &=&
 U_{nls:sit}(\la _1)|_{t = -\infty}
=
U_{nls}(\la _1)|_{t = -\infty}=U_{sit}(\la _1)|_{t 
= -\infty}=2i\sigma^3 \la _1,
    \nonumber \\  \sigma^3
{\rm v}_{nls:sit}^\infty &=& 
V_{nls:sit}(\la)|_{t = -\infty}=(V_{nls}(\la)+
V_{nls}(\la))|_{t = -\infty}=i\sigma^3 (2 \la_1^2+ (\la_1 -
w_{0})^{-1}N_0(z)) , \ll{UVinf}\eea
  which at the discrete 
 spectral parameter with complex value: $\la _1=k+i\eta, $
   take the explicit form
\be {\rm u}_{nls:sit}^\infty=i \la _1, \
 {\rm v}_{nls:sit}^\infty= {\rm v}_{nls}^{ \infty}
+{\rm v}_{sit}^{ \infty}, \ {\rm  v}_{nls}^{ \infty}=2i\la_1^2, \
{\rm v}_{sit}^{ \infty}=i(\la_1 - w_{0})^{-1}
N_0(z).\ll{uvINF} \ee  
 Inserting the needed complex valued expressions
 (\ref{uvINF}) in (\ref{Esol1})
 and grouping its real (Re) and
imaginary (Im) parts we  get the 1-soliton solution for the optical field 
 in  NLS-SIT equations 
(\ref{pnls},\ref{pN}) in the familiar $sech $- form
\be E= -2i\eta \  {\rm sech} 2 \zeta e^{2i\theta},
 \ \zeta= \eta(t-t_0-v z), \ \theta= \omega z+kt+\phi_0 ,\ll{solit}\ee 
where   $t_0$ and    $ \phi_0 $  are constant time  and  phase shift.   
 {\it Inverse speed } $ v$  and {\it phase rotation} $\omega$  
for the NLS-SIT soliton (\ref{solit}) are given by the {\it superposition} 
\be
 v=v_{nls}+ v_{sit}, 
    \ \omega= \omega_{nls}+  \omega_{sit}, 
 \ll{vom} \ee
of the corresponding parameters 
from   the NLS and the SIT subsystems, derived from (\ref{uvINF}) using $\la
_1=k+i\eta $  
as \bea
 v_{nls}=-\frac 1 \eta {Im[ {\rm v}_{nls}^{ \infty}]} =-4k, \ 
\omega_{nls}=Re[ \tilde{\rm v}_{nls}^{ \infty}]=2(k^2-\eta^2), \ll{vwnls} \\
   v_{sit}=-  \frac  1 \eta Im [\tilde{\rm v}_{sit}^{ \infty}]   = \frac 1 
\rho    { f(z)} 
,  \ \omega_{sit}= Re [\tilde {\rm v}_{sit}^{ \infty}]  = \frac 
{\tilde k} \rho    { f(z)} , \ll{vwsit}\eea
 with $ \tilde k=k -w_0 , \  \rho= 
{\tilde k}^2+\eta ^2,    
  $ and $ f(z)=   \frac  1 z \int^z N_0(z') dz' $.

It is intriguing to note that, since the z-evolution of the optical field
$E$
 in the NLS-SIT model follows the superposition  rule (\ref{vom})
 contributed separately by the NLS and the SIT parts,  
the term   $iE_z $ in equation (\ref{pnls}), evolving 
according to  solution 
 (\ref{solit}),
breaks up into two parts: one follows
the NLS contribution with  parameters (\ref{vwnls})
 and satisfies the pure NLS part of the 
equation in the {\it left hand side}, while the other
 part equates to the SIT deformation $2p_1 $  in 
the  {\it right hand side} of (\ref{pnls}) involving the related 
 parameters (\ref{vwsit}).  Using this dynamics we derive the soliton solution 
 for the  dipole $p_1$ from  (\ref{solit}),
in   the 
form
\be p_1 =  \ \frac \eta  \rho \ N_0
 {\rm sech} 2 \zeta (i\tilde k- \eta {\rm tanh} 2 \zeta )e^{2i \theta},
\nonumber \\ 
\ll{solitp}\ee
with $\zeta, \theta $ as expressed in (\ref{solit}). 
Inserting  solutions (\ref{solit},\ref{solitp})
 for $E $ and $p_1$   in    (\ref{pN}) and
 integrating by $t $ we  derive further the solution for  population inversion 
\be
N_1= N_{0}(1-\frac {\eta^2} {\rho} \  {\rm sech}^2 2 \zeta )
, \ll{Nsolit}\ee
again in the solitonic form
with arbitrary function $N_0(z) =N_1 (t \to -\infty)$, adjusted by the
  integration {\it constant}.  
 We obtain thus the complete set of  exact soliton solutions
to    the  NLS-SIT equations 
(\ref{pnls}-\ref{pN})
as (\ref{solit}) for the optical 
field  $E $, (\ref{solitp}) for the dipole $p_1 $
 and  (\ref{Nsolit}) for the population inversion $N_1$.
A  beautiful
 interaction pattern  can be  noticed  in these   solutions, manifested  in  the 
 superposition  relations: $v=v_{nls}+ v_{sit}, 
    \ \omega= \omega_{nls}+  \omega_{sit}$,
  for the  solitonic parametersappearing in   (\ref{solit},\ref{vom}).
 Intriguingly, 
in the absence of the SIT system with  $p_1=N_1=0 $, when 
  the  coupled NLS-SIT equations  
 reduce    to the  NLS  equation (\ref{nls}) 
 for the  field $E$, one recovers from     (\ref{solit})
 the  
well known NLS soliton
 by   simply  putting  $ v_{sit}=\omega_{sit}=0$
 due to  the vanishing
of (\ref{Vsit}). Therefore, the NLS soliton takes
 exactly  the same form as  (\ref{solit}),      
 though the  parameters are reduced  to  pure NLS case: 
 $v= v_{nls}, \ \omega=
\omega_{nls}$.
Similarly,   we can directly get  the   soliton   solution 
for the
pure SIT equations (\ref{SIT})
   in  the same form
(\ref{solit},\ref{solitp},\ref{Nsolit}),    but with  
   soliton parameters  reducing to $v= v_{sit}, \
 \omega=\omega_{sit}$, due to  switching off the 
NLS influence: $v_{nls}= \omega_{nls}=0 $.
Thus
   our exact  NLS-SIT soliton   
can  reproduce  the solutions for both the   NLS and the SIT equations
in a unified way, consistent with   the      ISM.
  However this rich interaction picture seems to have been 
missed   
  in a  well known earlier work \c{nakazawa1,nakazawa2}, leading to  
  wrong
conclusions in the general case.  In  particular, the soliton solution 
for the NLS-SIT equation  presented  in  \c{nakazawa1,nakazawa2} 
 gives the  expression 
 for  the {\it pulse delay}  as $\delta=\frac n c (1+\gamma)$ 
((4.9) in \c{nakazawa2}),
 which  is
equivalent to  the inverse soliton speed for the SIT ((2.22)
in   \c{nakazawa2}), i.e. \ 
 $v \equiv \delta=v_{sit} $, in our notation  \c{solNLSsit}. Similarly, 
 the {\it phase rotation} in 
 \c{nakazawa1,nakazawa2}  is given as $\alpha=2\eta^2 $, meaning
 $\omega \equiv \alpha=-\omega_{nls}$,  (at $k=0 $, see (\ref{vwnls})) in our
notation
 \c{solNLSsit}. Both these results 
for the coupled NLS-SIT equations appear to be  incomplete,  when  compared  with 
 our exact result (\ref{vom}). It is clear that,
the solution  of \c{nakazawa1,nakazawa2} can be justified only in a very limited
sense, when $v_{nls}=\omega_{sit}=0 $ and  therefore  unlike our soliton solution
 can not interpolate between the 
 solutions  of the NLS and the SIT equations.

This partial result unfortunately led to  wrong conclusions, for
the NLS-SIT system  in general,     
  stating  that (sect. IV \c{nakazawa2}),
     {\it the normalized speed} (i.e. $ \delta^{-1} $) {\it
 of the  NLS-SIT soliton is  
 determined only by    the SIT  effect }
 and similarly, {\it  the
z dependence of the phase  of the 
  dipole } (i.e. $ \alpha $)
{\it is  determined solely by the nonlinear phase change due 
   to  the NLS soliton}. Our exact   solutions
 for the  optical  field $E$
 (\ref{solit}) and the dipole  $p_1 $ (\ref{solitp}) 
 with  correct   expressions
 (\ref{vom}),   conclude on the other hand   that,
  only a part (i.e. $v_{sit} $) in the normalized speed
 $v^{-1}=(v_{nls}+v_{sit})^{-1}$ of  the NLS-SIT soliton is
 determined by the SIT effect, while there is  an additional
contribution 
 coming   from the NLS part $v_{nls} $.
   Similarly,
 the {z dependence  of the phase } of the dipole and the input optical 
 field gets  contribution 
from both the  NLS and the  SIT parts as 
$ \omega= \omega_{nls}+  \omega_{sit}
$, consistent with the interaction picture  in the coupled NLS-SIT system.
\\ \\
 {\bf III. CONTROLLING OPTICAL SOLITON EXPLOITING INTEGRABLE STRUCTURES}


Based on the integrable structures 
 underlying the NLS-SIT
system describing the propagation of optical soliton in fiber guided 
doped medium, we propose two  possible ways for controlling the
amplitude and width of the 
optical pulses.  
\\{\bf  A.  Soliton control by regulating initial population inversion}\\
It is  commonly believed that,  the exact
  soliton solution of a  homogeneous equation 
  always moves   with a constant speed, width and frequency,
 as in the case of the NLS
  soliton (\ref{vwnls})  with constant values for $v_{nls},\omega_{nls}$.
However, it is crucial to note  that,
   for the NLS-SIT soliton the parameters   $(v, \omega)$, 
  as evident from (\ref{vom},\ref{vwsit}) can become  
variable  functions,
depending on  the initial population inversion
 $N_0(z) $ (\ref{BC}). 
Due to this   peculiarity of    integrable structure of the NLS-SIT
system, hidden in the expressions like
(\ref{UV},\ref{Vsit},\ref{uvINF},\ref{vwsit}), the soliton 
  {\it speed}:  $v^{-1}$ and {\it width}: 
$(\eta v)^{-1}$, as defined
  from the soliton argument  $\zeta $  (\ref{solit}), 
  can be  variable and   linked to
 a controllable arbitrary function $N_0(z) $.

We  show that,  this important 
 observation embedded in the integrability of the NLS-SIT system
 can open up  a new  avenue for  controlling the optical soliton
 propagating through the doped medium, by regulating its initial population
 inversion profile $N_0(z) $. This fact 
 however remained unexplored in   earlier investigations 
\c{nakazawa1,nakazawa2,kakei,porsezian},    due to the restriction to a fixed
   initial profile
$N_0(z)=-1 $.  
Note that, at this    particular value giving   $ f(z)= -1$,
  our more
general result (\ref{vwsit}) reduces to
the simplified expressions obtained  earlier:  
\be 
   v_{sit}
= -\frac  1 { \rho }  
,  \ \omega_{sit}
=
- \frac 
{\tilde k} \rho ,  \ll{vwsit0} \ee

   The choice  for
the  initial population inversion 
  in the NLS-SIT model   as an arbitrary function $ N_0(z) > -1 $,  
that we propose here,
gives us the needed freedom for  obtaining   
   the excited  and the ground state  occupancies 
 at the initial moment as
   $|\tilde \nu |^2=\frac 1 2(1+ {N_0(z)}) $ 
  and   
   $|\nu |^2=\frac 1 2(1- {N_0(z)})$, respectively. 
Therefore, for  $N_0 > - 1$, giving $|\tilde \nu |^2> 0 $,
we can   prepare  
 the dopant  atoms  
   initially in an 
excited state  by   optical prepumping, resulting to  
the creation of a laser-active amplifying medium with  its
intensity determined by   $N_0$.
Note that, only in such a case when more active dopant atoms are in the
excited state, the optical soliton can gain net energy \c{OptCom}.
  
 In addition to the 
    soliton  pulse amplification, 
  variable initial  profile  $ N_0(z)> -1$,
 permitted by the  integrability of the
  NLS-SIT system,
 can   play  a crucial
role   in controlling
the shape and dynamics of the optical soliton.
  It is   possible, as we see below, 
to    address the  important problem of 
  pulse  broadening     
  by   regulating the initial profile 
 of the  dopant atoms.
 For example, 
a solitonic pulse governed by the NLS equation under 
  small  perturbation  by a   term 
$-i \frac   \Gamma 2  E $ with $\ \Gamma < < 1 $,
  would suffer 
broadening by a
factor $(4k\eta(z))^{-1}$, which
can be worked out through  the variational perturbation method as 
 $\eta(z)=\eta \ e^{-\Gamma z} $\c{agarwal}, which is 
 valid however  upto the range $\Gamma z \approx 1 $ along $z $.
 Beyond this
range with $z >> 1 $,  as shown by some other method,
the broadening of the pulse width 
 follows
 a different rule,  
 by increasing    linearly with  $z$ at a rate 
slower than the linear medium \c{agarwal}.
Though an 
 attenuation with intensity
 loss would also occur simultaneously, the broadening leads to more serious
problem of information loss and bandwidth limitation.
  Therefore 
  we  concentrate here only  on the
broadening problem  of the perturbed  NLS soliton,
 due to the  increasing
solitonic width $\frac 1 {4k\eta }\ e^{\Gamma z} $ along $z$, 
 as shown in Fig 1. As stated above for $z>> 1 $ it would follow a different
rule.  We show that, 
by transmitting  this solitonic pulse   through  a doped   resonant medium,
 described by an interacting  NLS-SIT model (\ref{pnls}-\ref{pN}),
it is possible 
 to control  the pulse broadening,  
 by   suitably     preparing  the initial 
 population inversion profile     $N_0(z)$.
 Fig 2a shows this controlling
effect, where the {\it broadening} of the solitonic pulse
suffered in Fig 1, is countered 
 by the {\it narrowing} of the pulse due to variable width
$V(z)=(v_{nls} + N_0(z) /( \rho \Gamma ) )^{-1} $,
 by taking $N_0(z) \sim \eta(z)^{-1}$.
Note that  the  profile $N_0(z) $  has to be adjusted   
differently at different ranges, as mentioned above,   to control
the   broadening  in  the respective regions for  a wide range  of $z$.
The soliton dynamics  
 would also  change to a  
 variable speed $V(z) $, 
 possible
due to  the energy 
 supplied by  optical prepumping.

This    potential   opportunity  for controlling the  pulse width,
hidden in the integrable property  of the NLS-SIT system, as explained above,
  was  missed in  earlier investigations     
 \c{nakazawa1,nakazawa2,kakei,porsezian}, 
since 
  the initial atoms are
 usually  taken     in their  ground state:
 $ \ |\nu|^2=1, \ |\tilde \nu|^2=0, $ by restricting   to   $N_0(z)=-1 $.
 
\begin{figure}[h!]
\qquad\qquad \qquad\qquad\qquad\qquad 
\includegraphics[width=5.3cm,height=5.3 cm]
{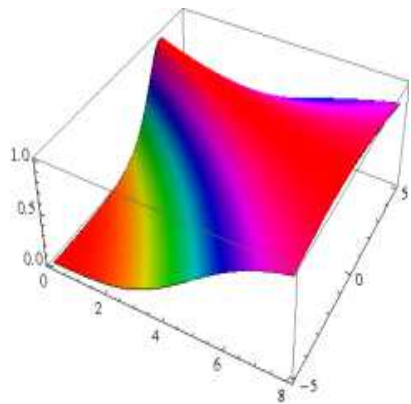}
\caption{ Broadening of the perturbed NLS soliton $|E(z,t)| $
 along the fiber,    moving with
a constant speed
with parameter  choice  $k= 0.25, \
\eta =0.50, \ \Gamma=0.28 $   }
\end{figure}

\begin{figure}[h!]
\begin{center}
\includegraphics[height=8.9 cm]
{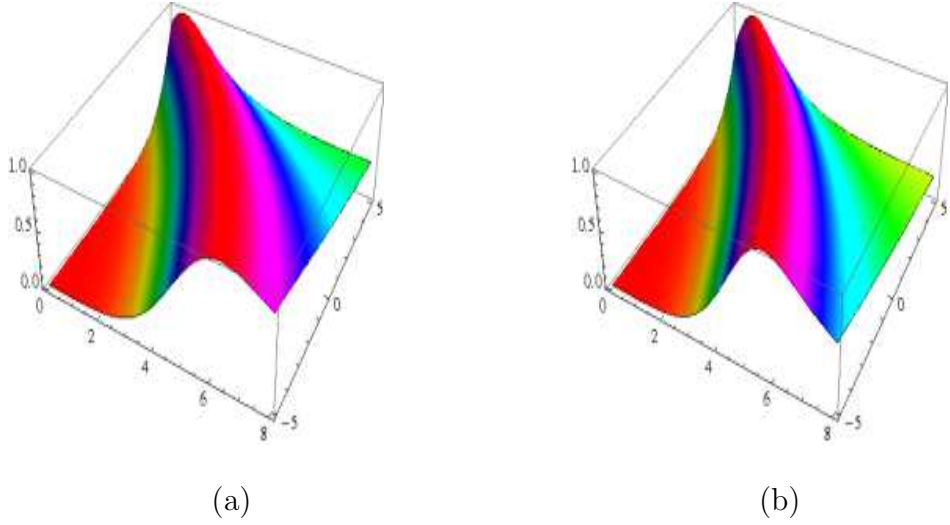}
\end{center}
\caption{a) Broadening  NLS  soliton pulse
is controlled by a coupled NLS-SIT system with  $ \ N_0(z)=
-0.11 e^{\Gamma z } $ and $w_0=0.3 $.   
  Variable  speed of the soliton  is 
  evident from its bending 
 in  the ($z,t$)- plane. b) Additional   control is achieved
 by  coupling to  a  second
 SIT system with   $N^{(2)}_0(z)= 0.11 e^{\Gamma  z}, \ $
  showing  an efficient restoration
of the  soliton  width.
 }
\end{figure}

\noindent {\bf B.  Enhanced soliton control through multiple doping}

Another promising opportunity   in managing  optical  soliton  in   fiber   
communication, emerging also from  the integrability of the  NLS-SIT 
model, is overlooked    completely in  earlier  investigations.
 This 
 is the   proposal of  enhancing the effect of
 amplification and control of 
  the optical soliton 
  by replacing the  conventional  single SIT system,
 the only case considered in the literature,  by a coupled
 multiple  SIT system,  using recursively the constrained integrable 
hierarchy   in  the  NLS-SIT model (see Fig. 3). 
 The physical meaning of coupling the  NLS equation to such  
 multi SIT system 
can be given through   a novel proposal of using   coupled   
   multiple doped
   resonant media, in place of a single doped medium.

For generating the governing  hierarchal   equations and showing
their  integrability,
 we   extend  Lax operator  $V(\la) $  (\ref{UV},\ref{Vsit}) by adding more 
deforming terms $V_{sitM}= i \sum_j^M  (\la-w_0)^{-j} G_j$, 
linked to the  $M $-th constrained    
 hierarchy  
 \c{kundu09} in   the 
  NLS-SIT system,  fixed at level  $M $ from
 the possible infinite sequence 
: $j=1,2, \ldots $.
In analogy with $G_1 $ 
(\ref{Vsit})  we can express the deforming matrices $G_j $, through
  dipole moment  $p_j $ and population inversion $N_j $
 of the $ j$-th doped resonant  medium.
For explicit  demonstration we restrict  to the next higher level $M=2 $
in the constrained hierarchy,   
by  considering only  an additional  SIT system  to the original 
NLS-SIT set. Compatibility 
of the Lax pair thus defined would generate an extended set of equations  given by  
 the same deformed NLS    (\ref{pnls}) coupled however to 
a  double SIT system
\bea
ip_{1t}&=&2(N_1 E-w_{0} p_1 -p_2), \ iN_{1t}=-(E p_1^*-E^* p_1), \nonumber \\ 
ip_{2t}&=&2 (N_2 E-w_{0} p_2 ), \ iN_{2t}=-(E p_2^*-E^* p_2), \ll{pNeM} \eea
 with
  induced polarization
 $p_2 $ and   population inversion   $ N_2$,
 linked to 
the additional doped medium described by  the 
second  SIT system.
We find intriguingly that, the exact soliton solution
 for the optical pulse $E $
 in this extended NLS-SIT
 model (\ref{pnls},\ref{pNeM}), 
 can be expressed   again  
 in   the same form (\ref{solit}), 
where  the soliton parameters  
are  to be modified with contributions from all its  interacting parts, i.e.
from the NLS as well as from the multiple  SIT system as
 $ \ v=v_{nls}+ v_{sit1} +v_{sit2} , \ \ 
     \omega= \omega_{nls}+  \omega_{sit1} +\omega_{sit2}. \ $
 Parameters $v_{nls}, \omega_{nls} $ and $  v_{sit1},  \omega_{sit1} $  have 
 the same expressions as found already in (\ref{vwnls},\ref{vwsit}),
 while the additional  SIT contribution $v_{sit2}, \omega_{sit2} $,
 can be derived following
 a similar argument as (\ref{vwsit}) in the form
 \bea
  v_{sit2}&=& - \frac 1 \eta Im [ ( \la _1-w_0 )^{-2}]  { f_2(z)} =
 2 \frac {\tilde k} {\rho^2}   { f_2(z)} 
, \nonumber \\  \omega_{sit2}&=& Re [ (\la _1-w_0 )^{-2}]     { f_2(z)}
 = 2 \frac {\tilde k ^2-\eta^2} {\rho^2}     { f_2(z)} ,\ll{vwsit2}\eea
with $ f_2(z)= \frac 1 z \int^z N_0^{(2)}(z') dz' ,$ involving an additional
arbitrary function $ N_0^{(2)}(z)=N_2(z, t=-\infty). $
 It opens
up  therefore  another novel way, hidden again in the integrable structure of the
NLS-SIT system, for an enhanced control of the soliton width and dynamics,
 by adding  a  coupled second 
SIT system, as shown in  Fig. 2b.

 This process of coupling the NLS equation to the set of multiple SIT
equations can be
continued within the framework of the integrable system, as mentioned above,
 creating a
form of directional  connected network  with feedback, as shown in
Fig. 3.
 In particular,
as evident from the coupled equations (\ref{pnls},\ref{pNeM}), the input
optical pulse $E$ would influence the  dipole field  $p_j $ and the   
  population inversion   $ N_j$ in all the resonant SIT media
 with $j=1,2, \ldots , M$,  while only
$p_1 $ from the first   medium gives feed back to the field $E $. On
the other hand, $p_{j+1} $ are coupled sequentially to $p_j $, across the
media, while
 $N_j $ are mutually coupled only  with $p_j $ from the same medium,   in
the  
 multiple SIT system with $j\in [1,M] $.
This network, would exhibit  more and more  manipulative power for
 control over width and amplification
  of the optical pulse,  enhanced  sequentially       
 by choosing  a set of initial condition
$N_0^{(j)}(z)=N_j(z,t=-\infty), j=1,2,\ldots M $  and
is based on the   notion of constrained hierarchy of the
integrable NLS-SIT system (see Fig. 3).

\begin{figure}[h!]
\begin{center}
\includegraphics[height=7.cm]
{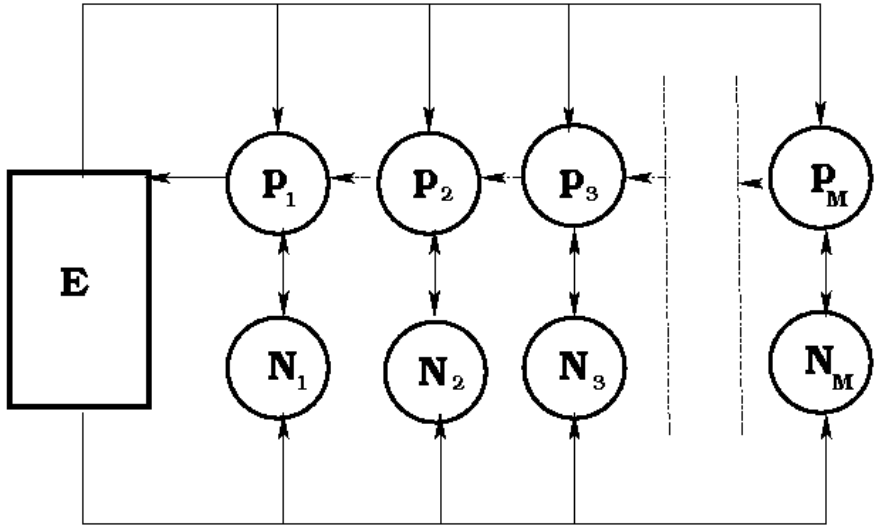}
\end{center}
\caption{ Connected  network of the NLS and the multiple   SIT system with
 $E$ as the   input optical
field, $p_j$ as the    induced polarization and $N_j $ as 
the population inversion of the $j $-th doped resonant
medium with $j=1,2,3,\ldots , M $.
The arrows show the directions of coupling with 
equations (\ref{pnls},\ref{pNeM})
 describing this network in the particular case of $M=2 $. Sequential
enhancement of the control of width and  amplification are    predicted by 
 this network, which  is
 consistent with the constrained
integrable hierarchy of the NLS-SIT system. }
\end{figure}

 This theoretical prediction, as presented schematically in Fig 3, 
 is an experimental challenge to incorporate
 the contribution of  coupled multiple SIT systems. Repeating the idea of
available experimental realization of single doped fiber medium,
  either  to a series 
of doped media coupled through induced polarization, or to multiple
doping  with  parallel coupling in a single medium,
 such experimental set up
is likely to be organized. 
\\ \\ \\   
\noindent {\bf IV. CONCLUDING REMARKS}

Exploring  the integrability  of the  coupled NLS-SIT system
 we have  given    novel proposals for     
 controlling   its solitonic pulse.
The  broadening problem of the optical pulse 
can be addressed  by adjusting
initial population inversion    of the dopant atoms,
linked  to the soliton width, by choosing 
  more general function  $N_0(z)> -1 $ for the initial profile, 
in place of  the  traditional restriction to $N_0(z)=-1$.
This  also allows  amplification of the signal  through initial excitation
by  prepumping  energy. 
The controlling effect can be refined further
by  using another integrable 
property of the coupled  NLS-SIT   model given by 
  its constrained   hierarchy.
The  idea is to replace the conventional single SIT system
by a network of sequentially coupled multiple  SIT media  with doping.
Each additional  SIT medium
 can  bring in a new tunable function $N^{(j)}_0(z)>-1, j=2,3, \ldots  $
   in the form
of  initial population inversion  of  additional dopant atoms, 
providing more manipulative  power for controlling the shape and dynamics 
 of the optical soliton.  
One set of dopant atoms 
 in the resonant medium is 
coupled to another set  
 by  induced  polarization,
 with all SIT media
  interacting in turn with the   input optical field.  
This network of interacting systems described by the 
 constrained hierarchy of the
integrable NLS-multiSIT equations is predicted to have   
  enhanced  control over solitonic  width and amplitude,     
which can increase  sequentially with the number of coupled SIT media.
In such a multi-doped media requiring higher  threshold  intensity
for the   formation of  solitonic pulse, one could   
possibly use   a multi-level  dopant
like   neodymium (Nd$^{3+}$), where
  with more than two available
levels
 the energy can be pumped throughout  the process, unlike in  two levels,
  resulting to a higher gain
\c{OptCom}.
 
Both of our theoretical    proposals with 
 applicable potentials can be worked out analytically 
in minute details through ISM, due to the   underlying integrability of the 
 system.


\begin{thebibliography}{99}

\bibitem{OptCom} G. P. Agarwal, {\it Fiber Optic Communication Systems},
(John Wiley, NY, 2002)
;

 V. Alwyn {\it Fiber Optic Technologies} (Cisco
Press, 2004); 

{Encyclopedia of Laser Physics \& Technology}, (Virtual
Web-Library, RP Photonics Consulting).
\bibitem{dismang} B. J. Eggleton et al, J. Lightwave Tech. {\bf 18}, 1418
(2000);

 X. F. Chen et al,  Photonics. Tech. Lett. IEEE {\bf 12 }, 1013 (2000); 

F. Poletti et al,  Photonics. Tech. Lett. IEEE {\bf 20 }, 1449
(2008). 
\bibitem{nlssolit}
L. F. Mollenauer et al, 
, R.H. Stolen and J. P. Gordon, 
Phys. Rev. Lett. {\bf
45}, 1095 (1980);

 A. Hasegawa and F. D. Tappert, Appl. Phys. Lett. 
{\bf
23}, 142 (1973);
   
 A. Hasegawa,  {\it Optical Fiber Solitons} (Springer,Berlin, 1989) 
\bibitem{agarwal} G. P. Agarwal, {\it Nonlinear Fiber Optics} (Acad. Press,
N.Y., 2007).
\bibitem{nlsdrab} F. M. Mitshke and L. F. Mollenauer, Opt. Lett. {\bf
11}, 657 (1986).
\bibitem{SIT} S. L. McCall and E. L. Hahn,  Phys. Rev. Lett. {\bf
18}, 908 (1967); Phys. Rev.  {\bf
183}, 457 (1969).
\bibitem{lamb} G. L. Lamb Jr., Rev. Mod. Phys. {\bf 43},  99 (1971).
\bibitem{Maimitsov} A. I. Maimistov and E. A. Manyakin, Sov. Phys. JETP {\bf
58},
 685 (1983).
 \bibitem{nakazawa1}  M. Nakazawa, E. Yamada and H. Kubota,
 Phys. Rev. Lett. {\bf
66}, 2625 (1991). \ 
\bibitem{nakazawa2}  M. Nakazawa, E. Yamada and H. Kubota,
 Phys. Rev.  {\bf
A 44}, 5973 (1991).
\bibitem{soliton}
 M. Ablowitz, D. J. Kaup, A. C. Newell and H. Segur,
 Stud. Appl. Math. {\bf 53}, 294   (1974);

M. Ablowitz and H. Segur, Solitons and Inverse Scattering Transforms
(SIAM, Philadelphia, 1981); 

 S. Novikov et al, Theory of Solitons (Consultants Bureau, NY, 1984).
\bibitem{kundu09}
 A. Kundu, J. Math Phys. {\bf 50}, 102702 (2009).

\bibitem{solNLSsit} Comparing  NLS-SIT  soliton (4.4) in \c{nakazawa2}
  with our  (\ref{solit})
 we  identify  pulse delay   $\delta $ with our $v$,
phase rotation $\alpha $ with our $\omega $,  inverse soliton speed
 $\frac n c (1+\gamma)$  
 with our $v_{sit}$ and $2 \eta ^2 $ with our $\omega_{sit} $ at $k=0 $.
\bibitem{kakei} S. Kakei and J Satsuma, J. Phys. Soc. Jpn. {\bf 63}, 885
(1994). 
\bibitem{porsezian} K. Porsezian and K. Nakkeeran, Phys. Rev. Lett. {\bf
74}, 2941 (1995). 

\end{thebibliography}
 \end{document}